\documentclass[12pt]{iopart}

\usepackage[dvips]{graphicx}
\usepackage{array}
\usepackage{amssymb}
\usepackage{hhline}
\usepackage{longtable}
\usepackage{dcolumn}
\usepackage{bm}
\usepackage{subfigure}
\usepackage[usenames]{color}
\begin{document}

\title[Random walks in small-world exponential treelike networks]{Random walks in small-world exponential treelike networks}

\author{Zhongzhi Zhang$^1$${}^,$$^2$, Xintong Li$^1$${}^,$$^2$, Yuan Lin$^1$${}^,$$^2$,  Guanrong Chen$^3$}

\address{$^1$ School of Computer Science, Fudan
University, Shanghai 200433, China}
\address{$^2$ Shanghai Key Lab of
Intelligent Information Processing, Fudan University, Shanghai
200433, China}
\address{$^3$ Department of Electronic
Engineering, City University of Hong Kong, Hong Kong, China}
 \eads{\mailto{zhangzz@fudan.edu.cn},\mailto{eegchen@cityu.edu.hk}}

\begin{abstract}
In this paper, we investigate random walks in a
family of small-world trees having an exponential degree
distribution. First, we address a trapping problem, that is, a particular case of random walks with an immobile trap located at the initial node. We obtain
the exact mean trapping time defined as the average of first-passage
time (FPT) from all nodes to the trap, which scales linearly with
the network order $N$ in large networks. Then, we determine
analytically the mean sending time, which is the mean of the FPTs
from the initial node to all other nodes, and show that it grows
with $N$ in the order of $N \ln N$. After that, we compute the
precise global mean first-passage time among all pairs of nodes and
find that it also varies in the order of $N \ln N$ in the large
limit of $N$. After obtaining the relevant quantities, we compare
them with each other and related our results to the efficiency for
information transmission by regarding the walker as an information
messenger. Finally, we compare our results with those previously
reported for other trees with different structural properties (e.g.,
degree distribution), such as the standard fractal trees and the
scale-free small-world trees, and show that the shortest path between a pair of nodes in a tree is responsible for the scaling of FPT between the two nodes.
\end{abstract}


\maketitle

\section{\label{sec01}Introduction}

As a powerful mathematical tool that can describe a large number of
real natural and manmade systems, complex networks have received
considerable interest from a wide range of scientific communities
recently~\cite{AlBa02,Ne03,BoLaMoChHw06}. During the last decade,
main endeavors were devoted to understanding the structural features
and dynamics of various networks~\cite{DoGoMe08}. In particular,
treelike networks have attracted renewed attention, because the
so-called border tree motifs are present in numerous real-life
systems and play a significant
role~\cite{ViroTrCo08,ShBuCoKiHaSt08}. The absence of loops in a
treelike network has a drastic influence on diverse dynamic
processes running on the network, e.g., the voter
model~\cite{CaLoBaCePa05} and naming game~\cite{DaBaBaLO06}.

Among a plethora of dynamics, random walks in trees have received
increasing attention in recent years, since the problem is related
to a wide range of research fields, such as physics~\cite{SiHo89},
biology~\cite{LoBl10}, and cognitive science~\cite{Fi87}. One of the
most important quantities of random walks is the first-passage time
(FPT)~\cite{Re01} defined as the expected time for a walker starting
from a source point to first arrive at a target
node~\cite{CoBeMo05,CoBeTeVoKl07}, which encodes much information
about random-walk dynamics. Thus far, random walks in trees with different
structures have been intensively studied, including the standard
fractal trees (e.g., the
$T-$fractal~\cite{KaRe89,Ag08,HaRo08,LiWuZh10,ZhLiZhWuGu09} and the
Vicsek fractal~\cite{Vi83,ZhWuZhZhGuWa10}) and scale-free
trees~\cite{Bobe05,BaCaPa08,ZhLiMa11}. These works uncovered how the
mean first-passage time (MFPT), i.e., the average of FPTs between
some given pairs of nodes, scales with the network size (number of
nodes), and was thus helpful for understanding the impact of
structural properties on the behavior of MFPT. For example, it was shown that the MFPTs between two nodes in different trees behave different. But the main reason for the difference remains not well understood. On the other hand, in contrast to
the scale-free behavior~\cite{BaAl99}, some real networks display an
exponential distribution as well~\cite{AmScBaSt00}. Relevant work on
random walks on such networks is much less. Particularly, what is the main factor affecting the speed of diffusion in general trees is still not well understood. A goal of this work is to answer this question, at least partially.

In this paper, we study a simple random walk~\cite{NoRi04} on a
family of deterministically growing small-world trees exhibiting
an exponential form of degree distribution~\cite{BaCoDaFi09}. We
first address a trapping problem, which is a particular random walk
with a single trap positioned at the initially created node of the
networks. We derive analytically the mean tapping time (MTT) defined
as the average of the FPTs from all nodes to the trap, which varies
lineally with the network size $N$. We then investigate the partial mean
first-passage time (PMFPT) (i.e., the average of the FPTs from the initial
node to a randomly selected target node) and the global mean
first-passage time (GMFPT) that is the average of the FPTs among all
pairs of nodes in the networks. Both PMFPT and GMFPT are determined
through the connection between random walks and electrical networks.
In contrast to the MTT, both PMFPT and GMFPT  are asymptotic to $N \ln
N$ for large networks. We relate our results to the efficiency of
information diffusion by considering the walker as an information
messenger. We also compare our results with those found for other
trees with different architectures and consequently give possible
reasons for the behavioral difference of random walks between the
considered trees and other comparable trees.

\begin{figure}
\begin{center}
\includegraphics[width=0.6\linewidth,trim=0 0 0 0]{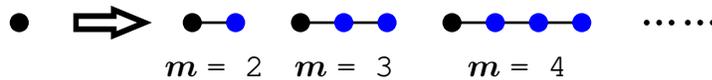}
\caption{(Color online) Construction method of the networks. Each existing node (in black) creates a path of $m$ nodes including the mother node itself.}
\label{Build}
\end{center}
\end{figure}

\section{\label{sec02}Network model and its properties}

\begin{figure}
\begin{center}
\includegraphics[width=0.6\linewidth,trim=0 10 0 0]{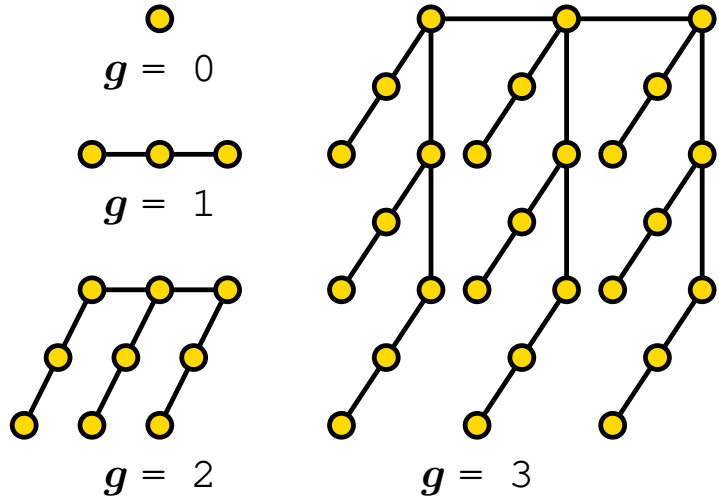}
\end{center}
\caption[kurzform]{(Color online) Illustration for the growing
process of a special network corresponding to $m=3$.}
\label{network}
\end{figure}

We first introduce the network model under consideration, which is
built iteratively and has a treelike structure~\cite{BaCoDaFi09}.
Let $U_{g}$ ($g\geq 0$) be the family of networks after $g$
iterations. Initially ($g=0$), $U_{0}$ is a single isolated node
without any edge, called the initial node below. For $g \geq 1$,
$U_{g}$ is obtained from $U_{g-1}$ by adding a path of $m$ nodes
($m$ is a natural number equal to or greater than 2) to each
existing node in $U_{g-1}$, see Fig.~\ref{Build}. By construction,
it is easy to know that the numbers of nodes and edges in $U_{g}$
are $N_g = m^g$ and $E_g=N_g-1=m^g-1$, respectively.
Figure~\ref{network} shows the growing process for a particular
network for the case of $m=3$. Notice that for $m=2$, the model
reduces to the deterministic uniform recursive tree proposed
in~\cite{JuKiKa02}, which has been extended and extensively studied
thereafter~\cite{DoMeOl06,BaCoDaFi08,QiZhDiZhGu09,ZhQiZhLiGu09,CoMi10}.

The networks considered here have some properties observed for many
real networks. According to the construction process, at any new
iteration, the degree of every old node increases by 1,
independently of the node degrees. Thus, the degree distribution of
the networks is exponential, instead of a power law~\cite{BaAl99}.
On the other hand, the diameter (i.e., the maximum of shortest-path distance between all pair of nodes) of $U_{g}$ is $(2g-1)(m-1)$ that
grows logarithmically with the network size, showing that the
networks are of small-world~\cite{WaSt98}. Moreover, some other
properties, e.g., the adjacency spectrum~\cite{BaCoDaFi09}, can also
be determined analytically.

\section{\label{sec03}Random walks on the networks}

After introducing the family $U_g$ of networks and their properties,
we will study the discrete random walks~\cite{NoRi04} performed on
$U_{g}$. At each time step, the walker (particle) jumps uniformly
(i.e., with the same probability) from its current position to any
of its neighboring nodes. One of the most important quantities
characterizing such a random walk is the FPT~\cite{Re01}. Let
$F_{i,j}(g)$ denote the FPT for a walker, staring from node $i$ in
$U_{g}$ to first arrive at node $j$. What we are concerned with is
how the scalings of FPTs behave as the network size increases.

In the following, we will focus on three cases of random walks.
Firstly, we will investigate a trapping issue, namely, random walks
with a single immobile trap located at the initial node, and
determine the MFPT to the initial node averaged over all nodes in
$U_{g}$. Then, we will compute the MFPT from the initial node to
another node selected uniformly from all nodes in $U_{g}$. Finally,
we will determine the MFPT between all pairs of nodes in $U_{g}$.

\subsection{MFPT from all other nodes to the initial node}

\begin{figure}
\begin{center}
\includegraphics[width=.6\linewidth,trim=0 0 0 0]{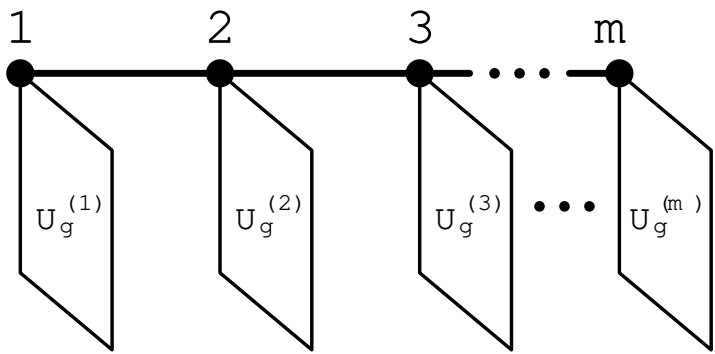}
\caption{Another construction of the network. The networks after
$g+1$ iterations, $U_{g+1}$, consist of $m$ replicas of $U_{g}$
denoted $U_g^{(1)}$, $U_g^{(2)}$, $U_g^{(3)}$, $\cdots$, and
$U_g^{(m)}$, which are connected to each other by adding $m-1$ edges
among the $m$ copies of the initial node.} \label{Joining}
\end{center}
\end{figure}

First, we study a particular trapping problem on $U_{g}$, in which
the single trap is positioned at the initial node. 
To facilitate computation, we introduce an
alterative construction method for the networks, which highlights
their self-similar architecture, as follows. Suppose one has $U_g$.
The next iteration of the network, $U_{g+1}$, can be obtained by
joining $m$ copies of $U_g$ in a way as illustrated by
Fig.~\ref{Joining}.

For convenience of description, we label the initial node in $U_{g}$
as $1$, while the duplicates of initial nodes in
$U_{g-1}^{(\eta)}(\eta=2,3,\ldots,m)$ are sequentially labeled as
$2$, $3$, $\ldots$, $m$. In addition, let $T_i(g)$ be the FPT,
$F_{i,1}(g)$, also called the trapping time, of node $i$ in $U_g$,
which is the expected time for a walker starting from $i$ to first
visit the trap node. Obviously, for all $g \geq 0$, $T_{1}(g)=0$.
Let $T_{\rm tot}(g)$ express the total trapping times for all nodes
in $U_g$, i.e.,
\begin{equation}\label{Trap01}
T_{\rm tot}(g)=\sum_{i \in U_{g}} T_i(g)\,.
\end{equation}
Then, the MFPT, also called the mean trapping time (MTT) and denoted
by $\langle T \rangle_g$, is the average of $T_i(g)$ over all
starting nodes distributed uniformly in $U_g$, given by
\begin{equation}\label{Trap02}
\langle T \rangle_g=\frac{1}{N_g}\sum_{i \in U_g}
T_i(g)=\frac{T_{\rm tot}(g)}{N_g}\,.
\end{equation}
Thus, to obtain $\langle T \rangle_g$, one should first determine
$T_{\rm tot}(g)$.


We first compute the FPTs, $F_{i+1,i}$ ($i=1,2, \ldots, m-1$)
between two arbitrary copies of the initial node of $U_{g-1}$ that
are directly connected to each other. Since $U_{g}$ has a treelike
structure, according to the result obtained previously
in~\cite{LiWuZh10,NoRi04a,NoKi06}, e.g., Eq.~(2) in~\cite{LiWuZh10},
we have
\begin{equation}\label{Trap03}
F_{i+1,i}(g)=2(m-i)m^{g-1}-1\,
\end{equation}
for all $i \in [1,m-1]$, yielding
\begin{eqnarray}\label{Trap04}
T_{\rm tot}(g)&=&T_{\rm tot}(g-1)+\sum_{\eta=2}^{m}\sum_{i \in
U_{g-1}^{(\eta)}} T_i(g)\nonumber \\
&=&mT_{\rm tot}(g-1)+\sum_{i=2}^{m} [N_{g-1}F_{i,1}(g)]\,.
\end{eqnarray}

Considering the treelike structure of $U_g$, we have
\begin{equation}\label{Trap05}
F_{i,1}(g)=F_{i,i-1}(g)+F_{i-1,i-2}(g)+\cdots +F_{3,2}(g)+F_{2,1}(g)\,.
\end{equation}
Plugging Eqs.~(\ref{Trap03}) and~(\ref{Trap05}) into
Eq.~(\ref{Trap04}), we obtain
\begin{equation}\label{Trap06}
T_{\rm tot}(g)=mT_{\rm
tot}(g-1)+\frac{1}{3}(2m^2-3m+1)m^{2g-1}-\frac{1}{2}(m-1)m^g\,.
\end{equation}
With the initial condition $T_{\rm tot}(0)=0$, Eq.~(\ref{Trap06}) is
inductively solved, giving
\begin{equation}\label{Trap07}
T_{\rm
tot}(g)=m^{2g}\left(\frac{2}{3}m-\frac{1}{3}\right)-m^g\left[\left(\frac{g}{2}+\frac{2}{3}\right)m-\left(\frac{g}{2}+\frac{1}{3}\right)\right].
\end{equation}

Inserting Eq.~(\ref{Trap07}) into Eq.~(\ref{Trap02}), we obtain a
closed-form expression for the MTT on network $U_g$, as
\begin{equation}\label{Trap08}
\langle T
\rangle_g=m^{g}\left(\frac{2}{3}m-\frac{1}{3}\right)-\left(\frac{g}{2}+\frac{2}{3}\right)m+\left(\frac{g}{2}+\frac{1}{3}\right)\,.
\end{equation}
Recalling $N_g=m^{g}$, we have $g=\ln N_g/\ln m$. Thus, $\langle T
\rangle_g$ can be expressed in terms of the network size $N_g$ as
\begin{equation}\label{Trap09}
\langle T
\rangle_g=\left(\frac{2}{3}m-\frac{1}{3}\right)N_g-\frac{\ln
N_g}{2\ln m}(m-1)-\left(\frac{2}{3}m-\frac{1}{3}\right)\,.
\end{equation}
Therefore, for large networks (i.e., $N_g \rightarrow \infty$),
\begin{equation}\label{Trapping06}
\langle T \rangle_g \sim N_g\,,
\end{equation}
implying that the MTT $\langle T \rangle_g$ increases linearly with
the network size, independently of the degree of the initial node.
This linear scaling is in sharp contrast to that of random
exponential trees~\cite{BaCaPa08}, in which the MTT depends on the
degree of trapping node.

\subsection{MFPT from the initial node to all other nodes}

By definition, $F_{1,i}(g)$ denotes the FPT of the walker visiting node $i$ for the
first time, assuming that the walker started at the initial node in
$U_g$. Let $\langle H \rangle_g$ represent the mean value of $F_{1,i}(g)$
averaged over all target nodes $i$ in network $U_g$, called the partial mean first-passage
time (PMFPT). Then, $\langle H \rangle_g$ is given by
\begin{equation}\label{Hit01}
\langle H \rangle_g=\frac{1}{N_g}\sum_{i \in U_g} F_{1,i}(g)\,.
\end{equation}
Thus, the problem of finding $\langle H \rangle_g$ is reduced to
determining the sum $\sum_{i \in U_g} F_{1,i}(g)$, denoted
$H_{\rm{tot}}(g)$.

Unfortunately, the method for computing $T_{\rm{tot}}(g)$ is not
suitable for $H_{\rm{tot}}(g)$. So, we seek for a feasible technique to derive
$H_{\rm{tot}}(g)$. Below, we will apply the link between effective
resistance and the FPTs for random walks~\cite{ChRaRuSm89,Te91} to
calculate $H_{\rm{tot}}(g)$ analytically. For this purpose, we
replace each edge of $U_{g}$ by a unit resistor to obtain the
corresponding resistor networks. To do so, let $R_{i,j}(g)$
represent the effective resistance between two nodes $i$ and $j$ of
$U_{g}$. Then, we have~\cite{ChRaRuSm89,Te91}:
$F_{i,j}(g)+F_{j,i}(g)=2E_gR_{i,j}(g)$, which leads to
\begin{equation}\label{Hit02}
H_{\rm{tot}}(g)+T_{\rm{tot}}(g)=2E_g\sum_{i \in U_g} R_{1,i}(g)\,.
\end{equation}

Equation~(\ref{Hit02}) shows that if we have the sum $\sum_{i \in
U_g} R_{1,i}(g)$ on the right-hand side, then we can easily obtain
$H_{\rm{tot}}(g)$. Since, for any tree, the effective resistance
$R_{i,j}(g)$ is equal to the geodesic distance $d_{i,j}(g)$ between
$i$ and $j$, this makes it possible to determine the sum $\sum_{i
\in U_g} R_{1,i}(g)$. To this end, we introduce a new quantity
$d_g$, which is the sum of shortest distances between the initial
node 1 and all other nodes in $U_g$. By definition, we have
\begin{equation}\label{Hit03}
d_g = \sum_{i \in U_g}d_{1,i}(g)\,.
\end{equation}
Considering the self-similar network structure (see
Fig.~\ref{Joining}), we can easily obtain the recursion relation
\begin{equation}\label{Hit04}
d_{g}=m\,d_{g-1}+\frac{m(m-1)}{2}N_{g-1}\,.
\end{equation}
Using $d_1=m(m-1)/2$, Eq.~(\ref{Hit04}) is solved, giving
\begin{equation}\label{Hit05}
d_g=\frac{1}{2}g(m-1)m^{g}\,.
\end{equation}
Then, we have
\begin{eqnarray}\label{Hit06}
H_{\rm{tot}}(g)&=&2E_gd_g-T_{\rm{tot}}(g)\nonumber\\
&=&m^{2g}\left
[g(m-1)-\frac{2}{3}m+\frac{1}{3}\right]-m^g\left(\frac{m-1}{2}g+\frac{2}{3}m-\frac{1}{3}\right)
\end{eqnarray}
and
\begin{equation}\label{Hit07}
\langle H \rangle_g=m^{g}\left
[g(m-1)-\frac{2}{3}m+\frac{1}{3}\right]-\left(\frac{m-1}{2}g+\frac{2}{3}m-\frac{1}{3}\right)\,.
\end{equation}

Equation~(\ref{Hit07}) can be rewritten as a function of the network
size $N_g$, as
\begin{eqnarray}\label{Hit08}
\langle H \rangle_g&=&N_{g}\left [\frac{(m-1)\ln N_{g}}{\ln
m}-\frac{2}{3}m+\frac{1}{3}\right]-\left[\frac{(m-1)\ln N_{g}}{2\ln
m}+\frac{2}{3}m-\frac{1}{3}\right]\,.
\end{eqnarray}
Therefore, in the limit of the large network size $N_{g}$,
\begin{equation}\label{Hit09}
\langle H \rangle_g \sim  N_{g}\ln N_{g}\,.
\end{equation}

\subsection{MFPT between all node pairs}

In what follows, we will calculate the MFPT $\langle F
\rangle_g$ among all node pairs in $U_g$, commonly called the global
mean first-passage time (GMFPT)~\cite{TeBeVo09}. 
It should be noted that the GMFPT has been studied in~\cite{CoMi10}.
Here we will derive an equivalent result using an approach different
from but relatively easier than that in~\cite{CoMi10}.

By definition, $\langle F \rangle_g$ is given by
\begin{equation}\label{MFPT01}
\langle F \rangle_g=\frac{F_{\rm
tot}(g)}{N_g(N_g-1)}=\frac{1}{N_g(N_g-1)}\sum_{i\neq j}^{N_g}
\sum_{j=1}^{N_g}F_{i,j}(g)\,,
\end{equation}
where the sum
\begin{equation}\label{MFPT02}
F_{\rm tot}(g)=\sum_{i \neq j}^{N_g} \sum_{j=1}^{N_g}F_{i,j}(g)
\end{equation}
denotes the sum of FPTs among all pairs of nodes. Hence, all that is
left to find $\langle F \rangle_g$ is to determine $F_{\rm tot}(g)$.

According to the relation between FPTs and the effective resistance,
we have
\begin{equation}\label{MFPT03}
F_{\rm tot}(g)=E_g\,\sum_{i \neq j}^{N_g}
\sum_{j=1}^{N_g}d_{i,j}(g)\,.
\end{equation}
For brevity, we use $D_g$ to denote $\sum_{i \neq j}^{N_g}
\sum_{j=1}^{N_g}d_{i,j}(g)$, which is the total geodesic distance
among all pairs of nodes $U_g$.

Since $U_g$ can be obtained by the juxtaposition of $m$ copies of
$U_{g-1}$ (i.e., $U_{g-1}^{(1)}$, $U_{g-1}^{(2)}$, $\cdots$, and
$U_{g-1}^{(m)}$) at the edge nodes (replicas of the initial node in
$U_{g-1}$), $D_g$ can be recast as
\begin{equation}\label{MFPT04}
D_g = m\,D_{g-1} + \Delta_g\,,
\end{equation}
where $\Delta_g$ is the sum over all shortest paths whose endpoints
are not in the same copy of $U_{g-1}$.

Denote $\Delta_g^{\alpha,\beta}$ as the sum of all shortest paths
with endpoints in $U_{g-1}^{\alpha}$ and $U_{g-1}^{\beta}$,
respectively. According to the value of the distance between two
edge nodes in $U_{g-1}^{\alpha}$ and $U_{g-1}^{\beta}$, we can
partition the sum of path length $\Delta_g^{\alpha,\beta}$ into $m-1$
classes, denoted by $\Delta_g^{\alpha,\beta}(q)$ with
$q=1,2,\ldots,m-1$ being the distance between the two boundary nodes
in $U_{g-1}^{\alpha}$ and $U_{g-1}^{\beta}$. That is, $\Delta_g^{\alpha,\beta}=\sum_{q=1}^{m-1}\Delta_g^{\alpha,\beta}(q)$.  It is easy to see that the number of elements in class
$\Delta_g^{\alpha,\beta}(q)$ is $m-q$. On the other hand, any two
elements belonging to $\Delta_g^{\alpha,\beta}(q)$ have an
identical length of $4N_{g-1}d_{g-1}+2q(N_{g-1})^2$. Then, the total
crossing path length $\Delta_g$ can be expressed as
\begin{equation}\label{MFPT05}
D_g = m\,D_{g-1}
+\sum_{q=1}^{m-1}(m-q)[4N_{g-1}d_{g-1}+2q(N_{g-1})^2]\,.
\end{equation}
With the initial condition $D_1=m(m-1)(m+1)/3$, Eq.~(\ref{MFPT05})
is solved to yield
\begin{equation}\label{MFPT06}
D_g =\frac{1}{3}m^{g}\left[(3g-2)m^{g+1}-(3g-1)m^{g}+2m-1\right] \,.
\end{equation}

Using the above-obtained results, the expression for $\langle F
\rangle_g$ reads
\begin{eqnarray}\label{MFPT07}
\langle F \rangle_g&=&\frac{E_gD_g}{N_g(N_g-1)}=\frac{D_g}{N_g}\nonumber\\
&=&m^{g}\left
[g(m-1)-\frac{2}{3}m+\frac{1}{3}\right]+\frac{2}{3}m-\frac{1}{3},
\end{eqnarray}
which can be rewritten in terms of $N_g$ in the following form:
\begin{eqnarray}\label{MFPT08}
\langle F \rangle_g=N_{g}\left [\frac{(m-1)\ln N_{g}}{\ln
m}-\frac{2}{3}m+\frac{1}{3}\right]+\frac{2}{3}m-\frac{1}{3}\,,
\end{eqnarray}
consistent with the result previously obtained in~\cite{CoMi10}.

Equation~(\ref{MFPT08}) uncovers the explicit dependence relation of
$\langle F \rangle_g$ on the network size $N_g$ and the parameter
$m$. In the case of $N_g\rightarrow \infty$, we have the following
expression:
\begin{equation}\label{MFPT09}
\langle F \rangle_g \sim  N_g \ln N_g\,.
\end{equation}

\subsection{Analysis and comparison}

Our results can be related to the efficiency of information transmission. Notice that if we consider the walker in the random-walk dynamics as
an information messenger, then the $\langle T \rangle_g$ measures the efficiency of the initial node (as a receiver) in receiving information, while $\langle H \rangle_g$ shows how efficient
of the initial node is as a sender to transmit information to other
nodes, and $\langle F \rangle_g$ is the efficiency of information sending when the
sender is distributed with equal probability among all nodes.

The above results, provided in Eqs.~(\ref{Trapping06})
and~(\ref{Hit09}), show evidently that the dominant behaviors for
$\langle T \rangle_g$ and $\langle H \rangle_g$ are different. The
former follows $\langle T \rangle_g \sim N_g$, while the latter
obeys $\langle H \rangle_g \sim N_g\,\ln N_g$, greater than the
former. This means that the initial node is more efficient in
receiving information than sending information. On the other hand,
Equation~(\ref{MFPT09}) together with Eqs.~(\ref{Trapping06})
and~(\ref{Hit09}) means that although the efficiency of the initial
node in receiving information is higher than that of the average
over other nodes, its ability of sending information is similar to
the others. Equations~(\ref{Trapping06}),~(\ref{Hit09})
and~(\ref{MFPT09}) also show that the linear scaling of the
efficiency for the initial node measured by the MTT is not a
representative property of the networks, but the $N_g \ln N_g$
behavior for the initial node sending information is so.

Our obtained results can be compared with those previously reported for other treelike networks.
Equations~(\ref{Trapping06}) and~(\ref{MFPT09}) imply that the
position of the trap significantly affects the scalings of the MTT
for the trapping problem with a single trap. 
This behavior is similar to that in the
small-world scale-free trees~\cite{ZhLiMa11,JuKiKa02}, but is in
contrast to that of trapping in the
$T-$fractals~\cite{Ag08,LiWuZh10,ZhLiZhWuGu09} and the fractal
scale-free trees~\cite{ZhLiMa11}, where the MTT does not depend on
the trap location.
In contrast, as shown in Eqs.~(\ref{Hit09}) and~(\ref{MFPT09}),
the initial node has the same dominant scaling of the PMFPT as that of
the average of PMFPTs over all senders. This equality between PMFPT and
GMFPT among all node pairs has also been observed for other trees,
including the $T-$fractals~\cite{Ag08,LiWuZh10,ZhLiZhWuGu09} and
fractal scale-free trees~\cite{ZhLiMa11}. However, the scaling of
PMFPT for different trees may have different behaviors.

In addition, the leading asymptotic $N_g \ln N_g$ dependence of GMFPT with the
network size is also compared to the scalings found from other
treelike networks with different degree distributions. In the
standard fractal trees, such as the
$T$-fractals~\cite{Ag08,LiWuZh10,ZhLiZhWuGu09} and the Vicsek
fractals~\cite{ZhWuZhZhGuWa10}, the GMFPT $\langle F \rangle$
increases superlinearly with the network size $N$, which has also been
observed from the family of scale-free trees with
fractality~\cite{ZhLiMa11}. For star graphs, the GMFPT $\langle F
\rangle$ grows linearly with $N$~\cite{ZhWuZhZhGuWa10}; while for
linear chains, $\langle F \rangle$ scales as a square root of
$N$~\cite{ZhWuZhZhGuWa10}. However, for the class of scale-free
small-world trees~\cite{ZhLiMa11,JuKiKa02}, the GMFPT $\langle F
\rangle$ also changes with $N$ as $\langle F \rangle \sim N\ln N$, which
follows the same scaling as that of the exponential trees studied
here.

Finally, combining the present work and the previous studies, it can be seen
that random walks in trees display rich behaviors in the context of
the FPT. At first sight, degree distribution is perhaps the root
responsible for the rich phenomena. However, in~\cite{ZhLiMa11}, it
was shown that the FPT in fractal and non-fractal scale-free
networks may exhibit quite disparate scalings, meaning that degree
distribution alone cannot determine the FPT for random walks on
trees. We argue that the FPT on trees is determined by the
short-path length from the resource node to the target node, while
the impact of other structural properties is encoded in the
short-path length, since the FPT $F_{ij}$ from an arbitrary node $i$
to $j$ is actually related to the FPTs of those node pairs for two
directly connected nodes along the unique short-path direction to
the target node, i.e.,
$F_{ij}=F_{ii_1}+F_{i_1i_2}+F_{i_2i_3}+\cdots+F_{i_xi_y}+F_{i_yj}$
provided that $i-i_1-i_2-i_3-\cdots -i_x-i_y-j$ is the shortest path
from $i$ to $j$. For example, the FPT on the non-fractal treelike
scale-free networks~\cite{ZhLiMa11} and on the exponential networks
studied here display similar behaviors, since both types of networks
are of small world with the average distance increasing
logarithmically in the network size, in spite of that they have
distinct degree distributions. As another example, the FPT on some standard fractal trees (e.g. the $T$-fractals~\cite{LiWuZh10,ZhLiZhWuGu09} and the Vicsek
fractals~\cite{ZhWuZhZhGuWa10}) displays a superlinear dependence on the system size, which is also due to their average distance.

\section{Conclusions}

We have presented a detailed analysis of the simple random walks on
a class of treelike small-world networks exhibiting an exponential
degree distribution. We first investigated the trapping problem,
focusing on a peculiar case with the trap fixed at the initial node,
and obtained the exact solution to the MTT, the dominating scaling
of which varies lineally with the network size $N$. We then studied
the random walks staring from the initial node, and determined
analytically the PMFPT from the initial node to all other nodes, whose
dominant behavior scales with $N$ as $N \ln N$. Moreover, we
determined explicitly the GMFPT among all node pairs and showed that
the GMFPT also increases with $N$ approximately as $N \ln N$. We
finally related our results in terms of information transmission
by regarding the walker as an information messenger, and compared
them with those previously reported results for other treelike
networks with disparate topological properties. Our work provides
new and useful insight into random-walk dynamics running on treelike
networks, and could further deepen our understanding of random walks on a tree~\cite{TeBeVo09}.

\section*{Acknowledgment}

This work was supported by the National Natural Science Foundation
of China under Grant No. 61074119 
and the Hong Kong Research Grants
Council under the GRF Grant CityU 1117/10E.

\end{document}